\documentclass[11pt,a4paper,oneside]{article}
\usepackage {graphicx}

\begin{document}

\title{Multi-Dimensional Hash Chains and Application to Micropayment Schemes}
\author{Quan Son Nguyen\\
\\
Faculty of Information Technology\\
Hanoi University of Technology\\
Hanoi, Vietnam\\
e-mail: sonnq@tinhvan.com} \maketitle

\begin{abstract}
One-way hash chains have been used in many micropayment schemes due
to their simplicity and efficiency. In this paper we introduce the
notion of multi-dimensional hash chains, which is a new
generalization of traditional one-way hash chains. We show that this
construction has storage-computational complexity of $O(\log _2 N)$
per chain element, which is comparable with the best result reported
in recent literature. Based on multi-dimensional hash chains, we
then propose two cash-like micropayment schemes, which have a number
of advantages in terms of efficiency and security. We also point out
some possible improvements to PayWord and similar schemes by using
multi-dimensional hash chains.
\end{abstract}

\section{Introduction}

One-way hash chains are an important cryptographic primitive and
have been used as a building block of a variety of cryptographic
applications such as access control, one-time signature, electronic
payment, on-line auction, etc.

In particular, there are many micropayment schemes based on one-way
hash chains, including PayWord \cite{RS96}, NetCard \cite{AMS96},
micro-iKP \cite{HSW96} and others.

By definition, micropayments are electronic payments of low value.
Other schemes designed for payments of high value normally use a
digital signature to authenticate every payment made. Such an
approach is not suitable for micropayments because of high
computational cost and bank processing cost in comparison with the
value of payment.

The use of hash chains in micropayment schemes allows minimizing the
use of digital signature, whose computation is far slower than the
computation of a hash function (according to \cite{RS96}, hash
functions are about 100 times faster than RSA signature
verification, and about 10,000 times faster than RSA signature
generation). Moreover, because a whole hash chain is authenticated
by a single digital signature on the root of chain, successive
micropayments can be aggregated into a single larger payment, thus
reducing bank processing cost.

There are a variety of improvements to hash chains. For example, in
the PayTree payment scheme \cite{JY96}, Jutla and Yung generalized
the hash chain to a hash tree. This construction allows the customer
to use parts of a tree to pay different vendors. Recently,
researchers have proposed a number of improved hash chains, which
are more efficient in terms of computational overhead and storage
requirement \cite{CJ02,Jak02,Sel03,HJP03}.

This paper is organized as follows. In section \ref{MDHC} we
introduce the notion of multi-dimensional hash chains (MDHC for
short). We also analyze efficiency of this construction and show
that RSA modular exponentiations could be used as one-way hash
functions of a MDHC. Section \ref{CashLikeSchemes} describes two
cash-like micropayment schemes based on MDHC, which have a number of
advantages in terms of efficiency and security. In section
\ref{ImprovePayword} we also examine some possible improvements to
PayWord and similar schemes. Finally, section \ref{Conclusion}
concludes the paper.

\section{Multi-Dimensional Hash Chain}
\label{MDHC}

\subsection{Motivation}

The notion of MDHC originates from one-way hash chains and one-way
accumulators \cite{BM94}. Here we briefly describe these two
constructions.

A hash chain is generated by applying a hash function multiple
times. Suppose that we have a one-way hash function $y = h(x)$ and
some starting value $x_n$. A hash chain consists of values $x_0 ,x_1
,x_2 ,...,x_n$ where $x_i = h(x_{i + 1})$ for $i = 0,1,...,n - 1$.
The value $x_0 = h^n(x_n )$ is called the root of hash chain. The
figure below depicts a hash chain of size $n$:
\begin{figure}[htbp]
\centerline{
  \mbox{\includegraphics[width=2.90in,height=0.53in]{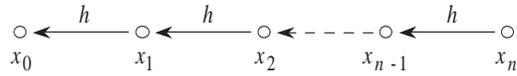}}
} \caption{A one-way hash chain} \label{HashChain}
\end{figure}

In contrast, a one-way accumulator is the output of multiple hash
functions, each of them applied only once:
\[
y = h_1 (h_2 (...(h_m (x))))
\]

In order to ensure that the output is uniquely determined regardless
of the application order, functions $h_1 ,h_2 ,...,h_m$ must be in
pairs commutative, i.e. $h_i (h_j (x)) = h_j (h_i (x))$ for any $x$.

Combining the two constructions described above, we define a
multi-dimensional hash chain as the result of multiple applications
of different commutative hash functions, so the root of an
$m$-dimensional hash chain is:
\[
X_0 = h_1 ^{n_1 }(h_2 ^{n_2}(...(h_m ^{n_m }(X_N ))))
\]

It is necessary to note that MDHC differs from other generalizations
of normal hash chain such as hash tree, which is used in PayTree
scheme. In particular such trees are generated from multiple leaf
nodes, while a MDHC is generated from a single starting value (i.e.
the value $X_N $ above).

\subsection{Definitions}

We begin with necessary definitions.

\noindent\textbf{\textit{Definition 1.}} Two functions $h_1 ,h_2
:\mbox{X} \to \mbox{X}$ are called commutative if $h_1 (h_2 (x)) =
h_2 (h_1 (x))$ for any $x \in \mbox{X}$.

\noindent\textbf{\textit{Definition 2.}} A one-way function
$h:\mbox{X} \to \mbox{Y}$ is called one-way independent of one-way
functions $h_1 ,h_2 ,...,h_m $ of the same domain if for any $x \in
\mbox{X}$, computing $h^{ - 1}(x)$ is intractable even if values
$h_1 ^{ - 1}(x),\;h_2 ^{ - 1}(x),...,\;h_m ^{ - 1}(x)$ are known.

And now we define MDHC as follows.

\noindent\textbf{\textit{Definition 3.}} Let $h_1 ,h_2 ,...,h_m $ be
$m$ one-way hash functions that are in pairs commutative and every
of them is one-way independent from all others. An $m$-dimensional
hash chain of size $(n_1 ,n_2 ,...,n_m )$ consists of values $x_{k_1
,k_2 ,...,k_m } $ where:
\[
x_{k_1 ,k_2 ,...,k_i ,...,k_m } = h_i (x_{k_1 ,k_2 ,...,k_i +
1,...,k_m }) \quad\textrm{for } i = 1,2,...,m \textrm{ and } k_i =
0,1,...,n_i
\]

The value $X_N = x_{n_1 ,n_2 ,...,n_m } $ is called the starting
node, and the value $X_0 = x_{0,0,...0} $ is called the root of the
MDHC, which is uniquely determined from $X_N $ due to commutativity
of hash functions:
\[
X_0 = h_1 ^{n_1 }(h_2 ^{n_2 }(...(h_m ^{n_m }(X_N )))) =
\prod\limits_{i = 1}^m {h_i ^{n_i }(X_N )}
\]

As an illustration, the figure below depicts a two-dimensional hash
chain of size (3,2):
\begin{figure}[htbp]
\centerline{
  \mbox{\includegraphics[width=2.71in,height=1.91in]{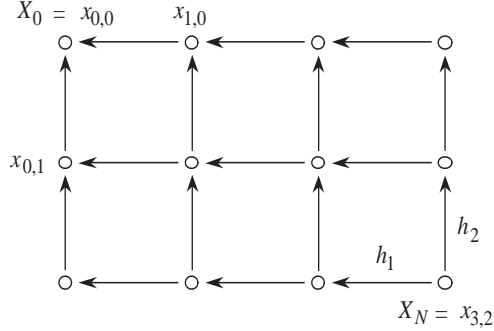}}
} \caption{A two-dimensional hash chain}
\label{TwoDimensionalHashChain}
\end{figure}

\subsection{Efficiency analysis}

In recent literature, there are a number of improvements to one-way
hash chains that aim to be more efficient in terms of computational
overhead and storage requirement. A widely used metric for one-way
hash chain efficiency is the storage-computational complexity, which
is the product of the traversal overhead and the storage required to
compute consecutive nodes of the hash chain.

It is easy to see that a linear hash chain size of $n$ has
storage-computational complexity of $O(n)$. In fact, if we
precompute and store all nodes (storage of $O(n))$, then no
computation is needed when a node is requested (traversal of
$O(1))$. Alternatively, we can store only the starting value, and
compute every node from the beginning each time it is requested.
This approach requires storage of $O(1)$ and $O(n)$ computations.
Also, if we store each of $t$ nodes, then storage of $O(n / t)$ and
$O(t)$ computations are required. So, in any case, the
storage-computational complexity of the linear hash chain is $O(n)$.

In \cite{CJ02,Jak02,Sel03} the authors have proposed new techniques
that make traversal and storage more efficient, which require
$O(\log _2 n)$ computations and $O(\log _2 n)$ storage, resulting in
storage-computational complexity of $O(\log _2 ^{\phantom{2}2}n)$.
Recently, Hu et al. \cite{HJP03} have presented a new hierarchical
construction for one-way hash chains that requires $O(\log _2 n)$
storage and only $O(1)$ traversal overhead.

In our case of $m$-dimensional hash chain of size $n$ (for
simplicity we assume all dimensions have the same size $n_1 = n_2 =
... = n_m = n)$, the number of nodes is $N = (n + 1)^m$. If we store
only the starting node of the chain (storage of $O(1))$ then maximal
number of calculations required to compute any node is $nm = n\log
_{n + 1} N$, or $\log _2 N$ if we select $n = 1$. In that case the
storage-computational complexity of MDHC is $O(\log _2 N)$, which is
equivalent to the results in \cite{HJP03}.

The advantage of MDHC is its simple implementation that does not
rely on the so-called pebbling technique, which is used in the
constructions mentioned above. However, the main limitation of this
construction is the fact that hash functions have to meet the
conditions described in the definition of MDHC. The RSA modular
exponentiation is known to meet these conditions, but it is not as
fast as the traditional hash functions, e.g. MD5 or SHA.

\subsection{RSA modular exponentiation}

Let consider the function of RSA modular exponentiation:
\[
y = x^c\bmod M
\]
\noindent where $c$ is some constant value and $M$ is an RSA
modulus, which is a product of two large primes of equal bit length
$p$ and $q$.

According to \cite{BM94}, the RSA modular exponentiation functions
with appropriately selected exponents could meet MDHC requirements.

First, obviously these functions are in pairs commutative: $h_i (h_j
(x)) = x^{c_i \,c_j }\bmod M = h_j (h_i (x))$

Second, one-wayness of these functions is derived from the RSA
assumption \cite{RSA78}, which states that the problem of finding
the modular root $x = y^{1 / c}\bmod M$ is intractable.

Finally, regarding one-way independence of functions, Shamir
\cite{Sha83} showed that if $c$ is not a divisor of the product
$c_{1\,} c_2 \,...\,c_m $ then the modular roots $y^{1 / c_1 }\bmod
M,\;y^{1 / c_2 }\bmod M,\;...,\;y^{1 / c_m }\bmod M$ are
insufficient to compute the value of $y^{1 / c}\bmod M$.

Therefore we can use the functions of RSA modular exponentiation as
one-way hash functions to construct multi-dimensional hash chains.

In that case we have following recursive expression:
\[
x_{k_1 ,k_2 ,...,k_i ,...,k_m } = (x_{k_1 ,k_2 ,...,k_i + 1,...,k_m
} )^{c_i }\bmod M \quad\textrm{for } i = 1,2,...,m \textrm{ and }
k_i = 0,1,...,n_i
\]
\noindent where $c_1 ,c_2 ,...,c_m $ are exponents of RSA functions
$h_1 ,h_2 ,...,h_m $ respectively.

Note that if one knows the factorization of $M$ (i.e. knows $p$ and
$q)$, then one can compute $X_0 $ quickly by using following
expression:
\[
X_0 = X_N \,^{\prod\limits_{i = 1}^m {c_i ^{n_i }\bmod E} }\bmod M
\]
\noindent where $E = \varphi (M) = (p - 1)(q - 1)$, and $\varphi $
denotes the Euler's totient function.

The expression above consists of only one modular exponentiation
with modulus $M$ and $\log _2 N$ modular multiplications with
modulus $E$. Since a multiplication is far faster than an
exponentiation, this expression allows us to compute $X_0 $ from
$X_N $ in a very effective manner.

\section{Cash-like Schemes Based on MDHC}
\label{CashLikeSchemes}

Cash-like payment schemes use the notion of electronic coin, which
is an authenticated (by the bank) bit string that is easy to verify,
but hard to forge. Examples of such coin are hash collisions (as in
MicroMint \cite{RS96}), or digital signatures (as in Ecash
\cite{Sch98}).

Let's recall the definition of MDHC. If we select the size of the
hash chain with $n = 1$ then all nodes $X_i = x_{0,0,...,1,...0} $
(with all $k_{j \ne i} = 0$, except $k_i = 1)$ have the same hash
value: $h_i (X_i ) = X_0 $. So we can use a pair $(X_i ,h_i )$ as an
electronic coin since:

\begin{itemize}\itemsep 0pt\parskip 0pt\vspace{-1ex}

\item[--] It is easy to verify by just one hashing.

\item[--] It is hard to forge because hash functions $h_i $ are
one-way, and their one-way independence assures that coin forgery is
impossible even if one knows other coins with the same root $X_0 $.

\end{itemize}\vspace{-1ex}

As a proof of that concept, we suggest two micropayment schemes
based on MDHC with the RSA modular exponentiation. We refer to these
as S1 and S2 schemes.

\subsection{The S1 scheme}

We assume that there are three parties involved in a micropayment
scheme, namely a bank (B), a customer (C) and a vendor (V). B is
trusted by both C and V.

\noindent\textsc{Setup}:

\begin{itemize}\itemsep 0pt\parskip 0pt\vspace{-1ex}

\item[--] B selects an RSA modulus $M = pq$ where $p$ and $q$ are large
\textit{safe} primes of equal bit length. A prime p is called safe
if $p = 2p' + 1$ where $p'$ is also an odd prime.

\item[--] B chooses $m$ constant values $c_1 ,c_2 ,...,c_m $ that satisfy
the condition of one-way independence, i.e. each $c_i $ is not a
factor of $\prod\nolimits_{j \ne i} {c_j } $. These values together
with modulus $M$ are public parameters and can be used for multiple
coin generations.

\item[--] To generate $m$ coins, B picks a random value
$X_N $ and computes:
\[
C = c_1 c_2 ...c_m \bmod E \quad\textrm{where } E = (p - 1)(q - 1)
\]
\[
X_0 = h_1 (h_2 (...(h_m (X_N )))) = X_N ^{\phantom{N}C}\bmod M
\]
\[
X_i = h_1 (h_2 (...(h_{i - 1} (h_{i + 1} (...(h_m (X_N ))))))) = X_N
^{\phantom{N}C\,c_i ^{ - 1}\bmod E}\bmod M\mbox{ , }i = 1,...,m
\]

Now B has $m$ coins $(X_i ,c_i )$.

\item[--] B keeps $X_0 $ in a public list of coin roots.

\item[--] For prevention of double-spending B keeps another list of all
unspent coins. In addition, B can also generate vendor-specific as
well as customer-specific coins by using some bit portions of
constants $c_i $ to form vendor ID and customer ID, similar to the
technique used in MicroMint scheme.

\item[--] C buys a sufficiently large number of coins from B before making
purchases.

\end{itemize}\vspace{-1ex}

\noindent\textsc{Payment}:

\begin{itemize}\itemsep 0pt\parskip 0pt\vspace{-1ex}

\item[--] C pays a coin $(X_i ,c_i )$ to vendor V.

\item[--] V verifies the coin by computing $X_0 = X_i ^{c_i }\bmod M$, and
checks if $X_0 $ is in the list of coin roots. Note that this list
is relative small and does not change frequently so C could keep it
locally.

\item[--] To assure that a coin was not double-spent, V either checks the
list of unspent coins on-line with B, or checks (off-line) the list
of coins he already received if the coin is vendor-specific.

\end{itemize}\vspace{-1ex}

\noindent\textsc{Redemption}:

\begin{itemize}\itemsep 0pt\parskip 0pt\vspace{-1ex}

\item[--] V deposits the coins he got from customers to B and receives an
amount corresponding to number of coins.

\end{itemize}\vspace{-1ex}

At the end of the coin validity period, C can sell unused coins back
to B or exchange them for new coins.

The proposed above scheme has several advantages:

\begin{itemize}\itemsep 0pt\parskip 0pt\vspace{-1ex}

\item[--] Coins are hard to forge under the RSA assumption.

\item[--] Payment can be made off-line by using vendor-specific coins.

\item[--] If customer-specific coins are not used, the scheme is anonymous
and untraceable because coins contain no customer information and
there are no links between coins.

\end{itemize}\vspace{-1ex}

However, the disadvantages of this scheme are:

\begin{itemize}\itemsep 0pt\parskip 0pt\vspace{-1ex}

\item[--] Generation and verification of coins is not very efficient. Each
coin requires one modular exponentiation to generate or verify it,
which is much slower than normal hash calculation.

\item[--] The list of unspent coins can be very big, though this is a
common problem of most coin-based schemes.

\end{itemize}\vspace{-1ex}

To overcome these disadvantages, we propose a modified scheme with
larger size hash chains (i.e. with $n > 1)$. In this scheme, B
generates $m$ chains of coins at once, rather than $m$ single coins.
Each coin chain is similar to the hash chain used in the PayWord
scheme.

\subsection{The S2 scheme}

\noindent\textsc{Setup}:

\begin{itemize}\itemsep 0pt\parskip 0pt\vspace{-1ex}

\item[--] B selects public parameters $M$ and $c_1 ,c_2 ,...,c_m $ in the
same way as in the S1 scheme. Let $n$ be the size of the hash chains
(for simplicity we assume all dimensions have the same size i.e.
$n_1 = n_2 = ... = n_m = n)$.

\item[--] B picks a random value $X_N $ and computes:
\[
C = c_1 ^nc_2 ^n...c_m ^n\bmod E \quad\textrm{where } E = (p - 1)(q
- 1)
\]
\[
X_0 = X_N ^{\phantom{N}C}\bmod M
\]
\[
X_i = X_N ^{\phantom{N}C\,c_i ^{ - n}\bmod E}\bmod M
\quad\textrm{for }i = 1,2,...,m
\]

Now B has $m$ coin chains $(X_i ,c_i )$. Each of those chains
contains exactly $n$ coins $(x_{i,j} ,c_i ,j)$ for $j = 1,2,...,n$
where:

\[
x_{i,j} = x_{i,j + 1} ^{\phantom{ij}c_i }\bmod M \quad\textrm{for }
i = 1,2,...,m \textrm{ and } j = 0,1,...,n - 1
\]
\[
x_{i,n} = X_i \quad\textrm{and}\quad x_{i,0} = X_0
\]

The coins from one coin chain must be paid to the same vendor.

\item[--] For double-spending prevention, now there is no need to keep
track of all unspent coins. Instead, B keeps the list of first coins
of all unused chains.

\item[--] As in the S1 scheme, coin chains can be vendor-specific as well
as customer-specific.

\item[--] C buys coin chains from B before making purchases.

\end{itemize}\vspace{-1ex}

\noindent\textsc{Payment}:

\begin{itemize}\itemsep 0pt\parskip 0pt\vspace{-1ex}

\item[--] C pays a vendor V the coins from a coin chain. The first coin of
the chain $(x_{i,1} ,c_i ,1)$ is verified by computing $X_0 =
x_{i,0} = x_{i,1} ^{\phantom{i}c_i }\bmod M$ and lookup of $X_0 $ in
the list of chain roots. It is also checked for double-spending by
lookup in the list of unused chains. Any subsequent coin is verified
by checking that it hashes to the previous coin in the chain, as in
the PayWord scheme:
\[
h_i (x_{i,j + 1} ) = x_{i,j + 1} ^{\phantom{ij}c_i }\bmod M \equiv
x_{i,j}
\]

\end{itemize}\vspace{-1ex}

\noindent\textsc{Redemption}:

\begin{itemize}\itemsep 0pt\parskip 0pt\vspace{-1ex}

\item[--] V deposits the last coin (i.e. the coin with highest index $j)$
of each coin chain he got from customers to B and receives an amount
corresponding to number of coins.

\end{itemize}\vspace{-1ex}

Comparing with the S1 scheme, this modified scheme retains all
advantages of S1, but storage requirement is reduced by factor of
$n$. In fact, B keeps track of only the first coins of $n$-coin
chains.

Another advantage of this scheme is more efficient coin generation.
Because B knows the factorization of M, he can compute the starting
node of a coin chain by just one modular exponentiation. Thus the
cost of this computational expensive operation is shared over all
coins of the chain. Similarly, B can also verify coin chains that he
got from vendors by computing one modular exponentiation per chain.

Generally speaking, the S2 scheme combines the advantages of two
approaches. A first approach uses unrelated coins that are
convenient for payments to multiple vendors. Another approach uses
chains of coins that are easy to generate and verify. In our scheme
different coin chains are unrelated, while coins within a chain are
generated and verified only by repeated hashing.

\section{Improve PayWord Scheme by Using MDHC}
\label{ImprovePayword}

The PayWord scheme has been proposed in \cite{RS96}. It is based on
one-way hash chains described in the section \ref{MDHC}. In this
scheme, before making purchases a customer C generates a hash chain
$x_0 ,x_1 ,...x_n $ (that is a chain of paywords) and sends his
signature of the root $x_0 $ to the vendor V. The customer then
makes a payment to V by revealing the next payword, which can be
verified by checking that it hashes to the previous payword.

The PayWord scheme allows a vendor to aggregate successive payments
from a customer by sending only last payword he got from the
customer to the bank for redemption. However, a vendor cannot
aggregate payments of different customers, nor can a customer use
the same chain of paywords to make payments to different vendors,
because there is no way to merge different hash chains.

By using MDHC, we can improve PayWord scheme in a number of ways.
Below we briefly describe two of such possible improvements. Note
that some irrelevant details in these descriptions are omitted for
convenience.

\subsection{Multiple denominations}

In the original PayWord scheme the size of the hash chain must be
large enough. For example, if each micropayment is worth 1 cent and
total payment is up to {\$}100, then a chain with size of 10,000
must be generated, which requires 10,000 hash calculations.

We can reduce the number of hash calculations by using MDHC instead
of linear hash chain. The idea is that every dimension of MDHC will
be associated with different weight (or denomination) according to
some number system (e.g. decimal or binary).

Suppose we have an $m$-dimensional hash chain with size of $n$. If
one step in the ($i$+1)$^{th}$ dimension is equivalent to $(n + 1)$
steps in $i^{th}$ dimension, then a node $x_{k_1 ,k_2 ,...,k_m } $
corresponds to the value:
\[
k_1 + k_2 (n + 1) + k_3 (n + 1)^2 + ... + k_m (n + 1)^{m - 1}
\]

The maximal value that could be represented by this hash chain is $N
= (n + 1)^m - 1$ and the number of hash calculations required to
generate the hash chain is $n\log _{n + 1} (N + 1)$. In the case of
a binary number system (i.e. $n = 1)$ it is $\log _2 (N + 1)$.

Returning to the example above, the hash chain now requires just 14
calculations to generate.

Similarly, verification of the payword also requires significantly
less calculations than in the case of the original PayWord scheme.

\subsection{Multiple vendors}

In the PayWord scheme a hash chain can be used for payments to only
one vendor. A customer must generate different hash chains for
payment to different vendors.

We can overcome this drawback by using MDHC as well. Let every
vendor $\mbox{V}_i $ in the payment system is assigned a different
hash function $h_i $ (i.e. a public parameter $c_i $ in the case of
RSA modular exponentiation).

Now, in order to make payment to $m$ different vendors, a customer
generates an $m$-dimensional hash chain with their public parameters
$c_i $ and signs its root. The customer then makes a payment to
$\mbox{V}_i $ by revealing the next payword in the $i^{th}$
dimension, starting from the root of hash chain.

In particular, if the current payword is $x_{k_1 ,k_2 ,...,k_i
,...,k_m } $, the next payword in $i^{th}$ dimension will be $x_{k_1
,k_2 ,...,k_i + 1,...,k_m } $.

At the end of the day, vendors deposit the last paywords they got to
the bank for redemption. The bank picks the last payword (which is
the one with highest indices) among paywords with certain root
(which all come from one customer). Finally, the bank credits
vendors $\mbox{V}_i $ by the amount equivalent to $k_i $, and debits
the customer's account accordingly.

There could be other possible improvements to the PayWord scheme by
using MDHC. For example we can aggregate payments of different
customers into a single MDHC that is generated by the bank, or we
can construct a payment scheme with multiple currencies, etc.

\section{Conclusion}
\label{Conclusion}

The proposed multi-dimensional hash chain is a simple and efficient
construction for one-way hash chains. Whereas a traditional one-way
hash chain has a storage-computational complexity of $O(n)$, our
construction achieves a complexity of $O(\log _2 n)$, which is
comparable with the best result among other recently proposed
constructions.

We show that multi-dimensional hash chains can be very useful in
micropayment schemes. In particular, we suggest two cash-like
micropayment schemes based on MDHC with RSA modular exponentiation
as one-way hash function. The first scheme utilizes coins that are
hard to forge under the RSA assumption. This scheme could be also
off-line and untraceable. The second scheme has additional
advantages including very efficient coin generation/verification and
much less storage requirements.

We also point out some possible improvements to PayWord and similar
schemes by using MDHC, including payword chains with multiple
denominations, and a scheme that allows payment to multiple vendors
using the same payword chain.

An open issue for our construction is whether another one-way hash
function can be found that meets MDHC requirements, and at the same
time is more efficient than RSA modular exponentiation.


\begin{thebibliography}{99}

\bibitem {AMS96}R. Anderson, H. Manifavas, and C. Sutherland. NetCard - a
practical electronic cash system. \textit{Proceedings of the 4th
Security Protocols International Workshop (Security Protocols)},
pp.49--57, Lecture Notes in Computer Science vol. 1189.
Springer-Verlag, Berlin, 1996.

\bibitem {BM94}J. Benaloh, M. de Mare. One-Way Accumulators: A Decentralized
Alternative to Digital Signatures. \textit{Advances in Cryptology --
EUROCRYPT `93}. LNCS, vol.765, pp.274--285, Springer-Verlag, 1994.

\bibitem {CJ02}D. Coppersmith and M. Jakobsson. Almost optimal hash sequence
traversal. \textit{Proceedings of the Fourth Conference on Financial
Cryptography (FC `02)}, Lecture Notes in Computer Science, 2002.

\bibitem {HJP03}Y. Hu, M. Jakobsson and A. Perrig. Efficient Constructions
for One-way Hash Chains. \textit{SCS Technical Report Collection}, 2003.\\
http://reports-archive.adm.cs.cmu.edu/anon/2003/CMU-CS-03-220.ps

\bibitem {HSW96}R. Hauser, M. Steiner, and M. Waidner. Micro-payments based
on iKP. \textit{Proceedings of the} \textit{14th Worldwide Congress
on Computer and Communications Security Protection}, pp.67--82,
Paris, 1996.

\bibitem {Jak02}M. Jakobsson. Fractal hash sequence representation and
traversal. \textit{Proceedings of the 2002 IEEE International
Symposium on Information Theory (ISIT `02)}, pp.437--444, 2002.

\bibitem {JY96}C. Jutla and M. Yung. PayTree: amortized-signature for
flexible micropayments. \textit{Proceedings of the 2nd USENIX
Workshop on Electronic Commerce}, pp.213--21, Oakland, California,
November 1996.

\bibitem {RS96}R. Rivest and A. Shamir. PayWord and MicroMint: two simple
micropayment schemes. \textit{Proceedings of the 4th Security
Protocols International Workshop (Security Protocols)}, pp.69--87,
Lecture Notes in Computer Science vol. 1189. Springer-Verlag,
Berlin, 1996.

\bibitem {RSA78}R. Rivest, A. Shamir, and L.M. Adleman. A method for
obtaining digital signatures and public-key cryptosystems. In
\textit{Communications of the ACM, 21(2).} pp.120--126, February
1978.

\bibitem {Sch98}B. Schoenmakers. Security aspects of the Ecash payment
system. \textit{State of the Art in Applied Cryptography: Course on
Computer Security and Industrial Cryptography - Revised Lectures},
pp.338--52, Lecture Notes in Computer Science vol. 1528.
Springer-Verlag, Berlin, 1998.

\bibitem {Sel03}Yaron Sella. On the computation-storage trade-offs of hash
chain traversal. \textit{Proceedings of Financial Cryptography 2003
(FC 2003)}, 2003.

\bibitem {Sha83}A. Shamir. On the Generation of Cryptographically Strong
Pseudorandom Sequences. \textit{ACM Transactions on Computer Systems
(TOCS)}, Volume 1, Issue 1, pp.38--44, 1983.

\end{thebibliography}
\end{document}